SHIKHA GUPTA
ANIMESH KUMAR

# SENTIMENT-AWARE ENHANCEMENTS OF PAGERANK-BASED CITATION METRIC, IMPACT FACTOR, AND H-INDEX FOR RANKING THE AUTHORS OF SCHOLARLY ARTICLES

**Abstract**

The paper has been accepted for publication in Computer Science journal: http://journals.agh.edu.pl/csci

Heretofore, the only way to evaluate an author has been frequency-based citation metrics that assume citations to be of a neutral sentiment. However, considering the sentiment behind citations aids in a better understanding of the viewpoints of fellow researchers for the scholarly output of an author.

We present sentiment-enhanced alternatives to three conventional metrics namely Impact Factor, H-index, and PageRank-based index. The proposal studies the impact of the proposed metrics on the ranking of authors.

We experimented with two datasets, collectively comprising almost 20,000 citation sentences. The evaluation of the proposed metrics revealed a signiEcant impact of sentiments on author ranking, evidenced by a weak Kendall coefEcient for the Author Impact Factor and H-index. However, the PageRank-based metric showed a moderate to strong correlation, due to its prestige-based attributes. Furthermore, a remarkable Rank-biased deviation exceeding 28% was seen in all cases, indicating a stronger rank deviation in top-ordered ranks.

# 1. Introduction

A research project begins with a concept, often culminating in a publication, typically in a conference or journal. The published article may mention other articles, mentions being called citations. While mentioning article is called the citing article, the mentioned article is called the cited article.

The study of the impact of authors and scholarly articles based on citation frequency is a popular area of research. There are two primary categorizations of citation-based approaches in the realm of evaluation of authors and articles. The first involves assessing the impact of an author/article as a function of citation counts. For instance, H-index is a well-known citation-frequency-based metric. Impact Factor is a metric that centers on total article count and total citation count values for an author/article. The second considers the prestige of the author/article. For instance, a citation metric based on the PageRank algorithm. It is assumed that each scholarly paper makes the same amount of scientific contribution and that each citation holds the same importance [34].

While the use of frequency-based indicators remains a common practice, it has been widely discussed that relying solely on the number of citations fails to truly measure the impact of the cited articles on the research domain [12, 19]. For instance, an article cited just to discuss its shortcomings and improvements [30] may not be considered at par with one cited for positive contribution.

In recent years, a tangential focus while evaluating authors/articles has been to consider the content of a citation with the intent to understand the meaning behind citations, leveraging the inherent nature of academic writing [14]. Indeed, the importance of a citation changes based on its impact on the citing work- enhancing the cited work or providing background knowledge [32].

## 1.1. Motivation for the Proposal

In the literature, the researchers have shown agreement against an over-reliance on pure quantitative metrics for measuring scientific impact [25]. In a recent study, Xu et al. [36] investigated the influence of qualitative aspects of a citation such as criticism or praise on research quality assessment by conducting a content analysis on top-rated papers from the Association for Computational Linguistics (ACL) conferences. The study concluded that the metric of citation polarity can help in better evaluation of the quality of research output and offer ideas for unbiased assessment of scholarly articles.

When examining citation motivation, negative citations are not solely indicative of criticism. Indeed, they frequently highlight the limitations and deficiencies within the referenced work. These shortcomings often serve as potential areas for improvement, suggesting that building upon the cited work may pave the way for enhancements.

However, it's crucial to acknowledge that while negative citations often signify limitations, their interpretation can be subjective and may not uniformly guarantee the inherent value or relevance of the cited material for the citing paper [36]. Considering citation polarity is, therefore, an important area of research. Taking into account the qualitative aspect of citations, specifically the sentiment's tone and polarity conveyed by the citing article, can facilitate a more precise assessment of an article's influence [36].

Even when there are several proposals for classifying citations according to their polarity and for computing the sentiment score of citations [1, 4, 5, 12, 14, 15, 17, 19, 20, 23, 25, 27, 29, 33, 38], only a few proposals [12, 16, 19, 21] address the problem of sentiment-aware citation metrics.

In addition, the literature is lacking in presenting sentiment-enhanced citation metrics for evaluation of authors of the scholarly articles and the current work offers a novel proposal.

### 1.2. Research Objectives

In the present study, we consider the characterization of citations on the basis of the expressed sentiment as positive, negative, or neutral. Usually, the terms "positive" and "negative" for citations involve discerning the polarity of the citing article's opinion towards the cited article for a specific citation instance (praising or criticizing a specific aspect) [26].

The present work contributes to the literature by addressing the following research objectives:
- Objective 1: Propose a method to compute the 'Aggregated Sentiment Score' for an author of a scholarly article.
- Objective 2: Propose sentiment-aware enhancements based on the authors' aggregated sentiment score for the well-known citation metrics (H-index, Impact Factor, PageRank-based metric) to evaluate author impact.
- Objective 3: Study the impact on the ranking of authors obtained by the proposed sentiment-aware metrics vs. frequency-based citation metrics.

### 1.3. Our Contributions

We aim to measure the impact of authors' contributions while relying on sentiment scores of the citation sentences. The set of citation sentences from citing articles is taken from the two datasets with pre-extracted citation contexts: Citation Sentiment Corpus [5] (Corpus 1) and Citation Intent Classification Dataset [9] (Corpus 2). The first dataset contains 8736 citations, and the second dataset contains 11020 citations, each along with their ground truth for sentiment.

After computing the ranks for authors and their articles using the proposed sentiment-based metrics, we conducted a comparison with their quantitative counterparts. Based on statistical analysis, our evaluation shows mostly weak but positive values for Kendall's Tau Coefficient when comparing the proposed metrics to traditional indices. It is worth noting that when measured using Rank-Biased Deviation (RBD), these deviations consistently exceeded 28%, highlighting the influence of the sentiment associated with citations. The statistical results are noted in the significant shifts in the author rankings for articles that have a strong sentiment, whether positive or negative. Remarkably, PageRank demonstrates moderate to high Kendall's Tau coefficient and a consistently low RBD among all indices, indicating that the inclusion of sentiment has a stronger impact when the metrics are based solely on popularity (frequency of citations).

In summary, the present proposal makes the following contributions:

- Considers sentiment polarity of the citations earned by the authors of scholarly articles in evaluating the author impact.
- Determines the sentiment score for each citation instance in a comprehensive collection of research articles. The experiments consider two datasets with almost 20,000 citation sentences.
- Develops a method for computing the aggregate sentiment score for the authors.
- Computes the overall sentiment score for each article and author in the collection.
- Proposes sentiment-aware alternatives to the well-known conventional citation metrics- H-index, Impact Factor, PageRank-based metric.
- Studies the impact of the proposed metrics on the ranking of the articles and the authors. The results are analyzed employing the following statistical measures: Kendall's Tau Rank Correlation and Rank Biased Distance (RBD).

### 1.4. Organization of the Article

In the rest of the article, we review the related work in Section 2. The datasets are described in Section 3. Section 4 presents the steps in the proposed methodology. Section 5 details the experiments and their results. In response to the first research objective, Section 5.1 details our approach for identifying the sentiment in citations for the authors and the articles. The proposed sentiment-aware ranking metrics (research objective 2) are described in Section 5.2. To address the third research objective, the influence of the proposed citation metrics is evaluated in Section 5.3. A discussion of the results is presented in Section 6. Section 7 concludes the article and suggests a direction for future work.

## 2. Related Work

While there's extensive research in the fields of Scientometrics, Sentiment analysis, and Network theory independently, only a limited number of studies exist at the intersection of the three domains, precisely where our work is positioned.

### 2.1. Sentiment Analysis of Citations

At the intersection of Sentiment analysis and Scientometrics is the sentiment analysis of citations, usually considered a two-step process, where the initial step identifies the explicit/implicit citation contexts and extracts their locations in the citing article.

The subsequent step of assigning sentiment to citation instances often employs standard classifiers. In recent years, there's been a notable rise in automated sentiment classification of citations [1, 4, 5, 12, 14, 15, 17, 19, 20, 23, 25, 27, 29, 33, 38].

2.1.1. Identifying Citation Contexts

In scholarly articles, it is typically observed that either a citation is explicitly indicated in a complete sentence (termed as a citation sentence) ending with a citation mark or may be implicit with the reference details in the cited work extending beyond the explicit citation sentence, encompassing nearby sentences without citation marks [39].

Several tools are available to extract citations from research articles based on the article's structure and citation style. In recent studies by Nazir et al. [26] and Aljuaid et al. [3], citation sentences were extracted after converting PDFs to text using XPDF and then parsing these text files with ParCit. ParCit, an openly accessible tool, recognizes various structural elements in research articles, such as titles, authors, and abstracts, aiding in the extraction of citation sentences.

2.1.2. Sentiment-based Classification of Citations

In the popular methodologies for sentiment-based classification of citations, various machine learning and lexicon-based approaches have been employed. Usually, before employing machine learning classifiers for sentiment analysis, text preprocessing methods are applied. The following examples showcase varied preprocessing strategies in sentiment analysis research. One of the studies [12] utilized positive and negative polarity words, along with part-of-speech and dependency tags for feature extraction. In contrast, another study [27] employed lemmatization, n-grams, stop words removal, and term-document frequency. Meanwhile, a different study [31] implemented term frequency-inverse document frequency (TF-IDF) and Word2Vec techniques for data preparation and representation.

The landscape of citation sentiment analysis has undergone significant evolution parallel to advancements in sentiment analysis and machine learning algorithms. Popular methodologies include Decision Trees, SVM, Naive Bayes, Decision Trees, Random Forest, and KNN [27]. Some proposals [38] incorporate sophisticated models such as CNN and BiLSTM, reflecting a notable paradigm shift in methodology and analysis techniques.

2.1.3. Challenges in Identifying Citation's Sentiment

Scientific texts present multiple hurdles for sentiment analysis, as sentiments are often concealed, written neutrally, and expressed through an objective style influenced by authors' biases [12]. The dual mode of writing, where criticism follows light appraisals, further complicates sentiment identification [22]. This complexity extends to other literary genres, also the variability in citation styles across publishers adds complexity to pinpointing citation locations as well as selecting the citation window size, especially when citations span multiple sentences [26].

A challenge encountered in sentiment-based classification of citations arises from significant bias within datasets, notably towards the neutral class. The bias towards the neutral class leads to class imbalance issues, causing erroneous predictions and potentially overfitting machine learning models [38]. Accurately detecting subtle sentiments, especially refined negative ones in citations, presents an ongoing challenge. This difficulty stems from their implicit nature and veiled criticisms, posing a significant hurdle for both algorithms and human perception [12, 19].

Another challenge arises from technical terms or specialized jargon in research areas, devoid of inherent sentiment, yet causing noise in sentiment score computation [37]. For example, the term 'Support' in 'Support Vector Machines' implies a positive sentiment, thereby complicating the sentiment analysis process.

Since the focus of the current work is on proposing sentiment-infused citation metrics for author ranking, we employ two corpora of pre-extracted citation instances. The popular lexical resource SentiWordNet [12, 19] is employed for deciding the sentiment scores of the citation sentences. Therefore, the proposed approach is not required to address the above challenges in its approach.

## 2.2. Assigning Sentiment Score to a Citation Instance

Apart from supervised machine learning algorithms, some lexicon-based resources have also evolved that provide sentiment scores for sentences followed by categorization using the scores. A citation with a "Positive" or a "Negative" polarity does not necessarily indicate "good" or "bad". While a citing article may criticize a specific aspect of a cited article, it may praise a different aspect of the same cited article. Therefore, sentiment scores are employed instead of simply classifying citations based on pure sentiment.

Many researchers have proposed the use of sentiment scores that are imbued with both sentiment and degree. Sentiment scores involve determining the sentiment of the citing paper towards the cited paper for a specific citation instance- appreciating a particular aspect or criticizing the same [12]. A study has employed SenticNet [37] that integrates concepts and semantics to offer polarity and semantic information for concepts. More popularly, SentiWordNet [12, 19] is employed as a lexical resource for identifying citation sentiment based on the generated sentiment score. SentiWordNet is a lexical resource that associates sets of cognitive synonyms (synsets) with sentiment scores (positive, negative, neutral).

## 2.3. Influence of Citation Sentiments on Scientific Articles

At the intersection of Scientometrics, Network theory, and Sentiment analysis lies research in citation content analysis offering a nuanced and qualitative perspective on a cited article [17]. In the past decade, the field of scientometrics has seen a rise in the thought that an article's impact or usefulness shouldn't solely rely on the number of times it's cited but also on the manner of its citation. This means that the opinions expressed by the citing authors can serve as a valuable gauge of an article's influence.

However, most of the existing bibliometric measures to evaluate scholarly articles and their authors are majorly quantitative with a focus on the number of times an article is cited, based on a prevailing assumption that research articles are generally cited in a positive manner[12]. Abu-Jbara et al. (2013) [1] critiqued conventional bibliometric measures for lacking the ability to distinguish between positive and negative citations. However, few studies can be found that consider sentiment when assessing the impact of scholarly articles [12, 16, 19, 21]. Moreover, we did not find any prior work that ranked and evaluated the authors based on the sentiment analysis of citations.

Kazi et al. [16] proposed to incorporate semantic similarity considering the same between citing and cited article along with the polarity of the text surrounding the citation sentence and self-citations. The authors employ SentiWordNet to generate sentiment scores and show that their results are parallel to those of the PageRank-based approaches and are an improvement over traditional citation counts.

Ma et al. [21] proposed a method to classify the citations by integrating the sentiment polarity with the data about the authors' reputation, encoded in the form of p-index. p-index is obtained by multiplying the h-index value by the number of positive citations raised by a value greater than 1 and number of negative citations raised by a fractional power. However, the authors do not give the rationale behind the formulation of the equation for computation of p-index. The authors concluded that integrating the reputation of the authors in the input data significantly improves the process of citation sentiment classification.

Kochhar and Ojha [19] proposed an equation-based sentiment-enhanced impact factor for articles, which incorporates multiple factors. The approach considers sentiment scores (computed using SentiWordNet), the impact factor of both the citing and cited authors, and the respective publishing journals. The authors concluded that it is important to consider the sentiment behind the citations instead of simply considering that each citation contributes equally to the evaluation of the article's impact.

Ghosh and Shah [12] highlighted the importance of considering sentiment of citation and presented ranking indexes to appraise the significance of research papers. Based on supervised ML (machine learning) classifier, the authors determine the polarity of a citation sentence. Thereafter, the classifier is used to assign sentiment scores to citation instances. Their sentiment-based metric to rank articles uses overall citation score obtained by summing up the citation scores from citing to cited papers. The authors also proposed that the PageRank-based approach to rank articles be modified by considering the associated sentiment while determining the edge weight. For any edge from citing to cited article, the score transferred is multiplied by the sentiment score of the citation instance.

We present a comparative analysis of articles focusing on sentiment-aware metrics in Table 1. This table offers insights into the reviewed literature, highlighting methodology, results and limitations among various studies in the field of sentiment-aware citation metrics.

We compare our work with the study by Ghosh and Shah [12]. As compared to others, the work by [12] comes nearest to our proposal. We mark the following differences:

- The focus of the present work is on proposing sentiment-aware citation metrics for ranking authors. Ghosh and Shah [12] have focused on ranking articles.
- Ghosh and Shah [12] identify the sentiment in each citation sentence using supervised machine learning classifier. The present approach preprocesses the citation sentences to make the data suitable for input to SentiWordNet. The sentiment scores are computed using the widely used SentiWordNet. The proposed technique precludes any bias that may arise from technical terms and class imbalance issues.
- The cited article may be mentioned more than once in the text of the citing article. For every pair of citing and cited author/article, we associate a composite score representing the citation frequency and sentiment score values. Moreover, the context of the cited article/author is aggregated and an "aggregate sentiment score" is assigned to the authors and articles. Ghosh and Shah [12] use the classifier to assign sentiment scores to citation instances.
- We propose sentiment-infused alternatives to the three well-known conventional frequency-based citation metrics for ranking authors- H-index, Impact Factor, PageRank-based metric. Ghosh and Shah [12] propose sentiment-based modifications for following metrics to rank articles- citation count and PageRank-based approach.
- The proposed approach incorporates in its PageRank-based approach, the aggregate sentiment score as the edge weight. On the other hand, Ghosh and Shah [12] proposed that the PageRank-based approach to rank articles be modified by associating the sentiment to the edge weight. The authors [12] mention that for any edge from citing to cited article, the score transferred is multiplied by the sentiment score of the citation instance.
- The proposed approach clearly defines the damping factor value appropriate to the domain for its PageRank-based approach. The value of damping factor is significant in the working of PageRank algorithm as it can induce notable deviations even with minor adjustments [7]. Conversely, Ghosh and Shah [12] do not discuss their selection of a damping factor value.

## 3. Data Description

We conducted sentiment analysis on scientific citations using two corpora [5, 9] comprised of citation sentences. As can be seen in Figure 1, the sentence containing the citation (for the "cited article") is called the "Citation Sentence". The article containing the citation sentence is referred to as the "citing article".

In tandem with our focus, we searched for datasets that provided pre-extracted citation sentences. The Citation Sentiment Corpus by Athar (2011) [5] is a widely employed dataset in the field of citation sentiment analysis and includes citation contexts extracted from both citing and cited papers within the AAN repository. This dataset (Corpus 1) is from the ACL Anthology, a repository renowned for housing research articles in computational linguistics. An instance in the dataset contains a single sentence. The dataset comprises 8736 citation instances. Figure 1 shows a sample from the dataset. The dataset contains the following information: Citation sentence, citing article ID, the cited article ID, and the sentiment of a citation.

| Article | Experiments | Results | Limitations |
|---|---|---|---|
| Ghosh & Shah, 2020 12 | 1. Extracted features from the dataset and employed a meta-classifier (DAGGING) to facilitate automatic annotation of an unlabeled corpus.<br>2. Above corpus was used for computing the ranks using the proposed indexes.<br>3. Evaluated the performance of the generated indexes by comparing each ranked list with others. The evaluation criterion involved identifying differences between the order of ranked lists | 1. The overall accuracy of the classifier came out to be 80.61%.<br>2. Inclusion of sentiment and link structure led to more than 25% of difference between two ranking indexes | 1. The methodology explains the training and application of the sentiment classifier but lacks details on the specific computation process for generating sentiment scores.<br>2. The study defines self-citations as instances where the Source paper and Target paper are the same. However, self-citation often extends to the author's previous works rather than being limited to the same paper.<br>3. The study does not explicitly state the chosen damping factor for the PageRank indices, which is a crucial parameter in PageRank algorithms.<br>4. Additionally, the research scope is limited to analyzing articles, neglecting exploration into authors or journals. |
| Kochhar & Ojha, 2020 19 | 1. Extracted citations from the AAN repository and preprocessed citation contexts for analysis.<br>2. Calculated sentiment scores for each citation instance using SentiWordNet.<br>3. Computed metric values based on a defined equation, incorporating publication and authors' impact factors | The rankings were determined by utilizing calculated impact factors for ten articles, with a primary emphasis on a broader set of 91 papers. This collection comprised the initial ten papers, along with the additional papers that cited them. | 1. Lack of detailed discussion regarding the handling of multiple citations between a pair of papers and self-citations.<br>2. The final results (ranks) were not systematically assessed for trends or deviations, limiting a comprehensive analysis of the experimental outcomes. A small dataset is considered.<br>3. The research scope is limited to analyzing articles, neglecting exploration into authors or journals. |
| Kazi et al., 2016 16 | 1. Computed sentiment scores using SentiWordNet, along with weighted self-citations and semantic similarity values.<br>2. Utilized the computed values to assign scores to citation edges in the network and applied a graph traversal method to calculate the citation index.<br>3. Evaluated the correlation between PageRank algorithm ranks and those obtained through the graph traversal approach. | 1. The Spearman coefficient between the Graph Traversal Method and Weighted PageRank yielded a highly significant value of 0.995. | 1. When assigning weights to various factors to compute edge weights, there is a comparatively lesser level of consideration.<br>2. Although Spearman is a commonly employed correlation metric, its robustness diminishes when computing correlation in scenarios with a substantial number of ties.<br>3. The research area diverges from ours as it focuses on an alternative method for computing PageRanks in citation networks. |
| Ma et al., 2016 21 | 1. Derived an index (p-index) by incorporating sentiment into the h-index through a dedicated equation.<br>2. Utilized the newly created p-index and additional author reputation information as extensions to the fundamental features for classification purposes.<br>3. Trained and tested an SVM model to assess the impact on accuracy resulting from the integration of the p-index and supplementary author reputation details. | 1. It was noted that the combined inclusion of Affiliation ID and Author ID yielded enhanced performance. | 1. The method for counting the number of positive and negative citations to compute the p-index lacks an explicit description.<br>2. The formulation of the equation for the p-index lacks thorough discussion and clarification.<br>3. Their research deviates from ours as they emphasize the utility of the derived index for enhancing the classification of citation sentiment. |

Table 1
Related Literature for the Primary Objective of Proposing Sentiment-enhanced Citation Metrics for Ranking of Authors

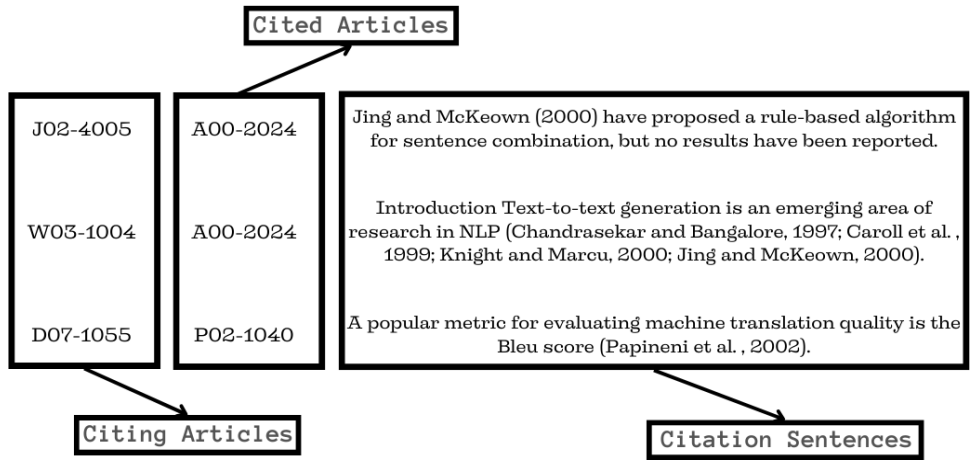

Figure 1. Examples of Citation Instances from Corpus 1

Comparison revealed that other available datasets have a lower number of citation instances. For example, Malakwani Ram's [23] dataset yielded approximately 5000 instances, Yousif's [38] approximately 2000 instances, and other existing datasets, such as Valenzuela's [32], contained around 500 instances.

The second dataset (Corpus 2) comprised 11,020 instances and was sourced from the Scicite intent corpus [9]. The corpus was created from a subset of the Semantic Scholar corpus [1]. Since it was initially designed for intent classification, we extracted the necessary fields, including citing article ID, cited article ID, and citation instances. The rationale for selecting this dataset was driven by its inherent benefit wherein utilizing article IDs enabled the efficient extraction of the author and other article details through the Semantic Scholar API. In addition, since the API supplies author IDs, author name disambiguation is not needed.

In addition, SciCite Intent Corpus (Corpus 2) demonstrated the impact of sentiment-aware metrics, particularly with a small number of citations. As shown in further sections, the sentiment-infused metrics serve as effective differentiators, functioning as tiebreakers, to efficiently distinguish between articles with minimal differences in citation counts. For instance, consider the citation sentence: "Our data also showed that lesions in the head and neck region responded better to PDL therapy than in other regions; similar results were reported by several studies [16, 24 â€" 28]." This sentence attributes a citation to the author identified by the author ID "10616537," who has received a singular citation, thereby placing the author at the shared rank of 11588 with several others, including author ID "3881197". However, an examination incorporating sentiment analysis yielded a score of "0.75" and "-1.5" and substantially different corresponding ranks of 2681 and 23267 respectively for the authors. A detailed comparative analysis is expounded upon in Section 6.

---

[1] https://semantirsrholar.org/

## 4. Proposed Method

The focus of the present work is on proposing sentiment-aware citation metrics for ranking authors. We use datasets with pre-extracted citation contexts. A flowchart to outline the steps involved in the proposed method is presented in Figure 2. Following are the broad steps that the proposed method follows:

- Step 1 corresponds to the first research objective (Section 5.1). We pre-process the citation sentences to make the data suitable for input to SentiWordNet. The sentiment scores are computed using the widely used SentiWordNet which prevents any bias that may arise from technical terms and class imbalance issues. We assign a composite score, which we call the "Aggregate Sentiment Score", to the authors and articles. This entailed identifying the unique authors in the dataset and removing self-citations before aggregating the scores for each author.
- Step 2 corresponds to the second research objective (Section 5.2). We propose sentiment-aware alternatives to the three well-known conventional frequency-based citation metrics- H-index, Impact Factor, PageRank-based metric. Each of the three proposed metrics replace use of simple citation frequency with the use of the composite sentiment score. To the best of our knowledge, this is first such proposal for ranking authors.
- In the end, corresponding to the third research objective (Section 5.3), we study the impact on author rankings of the proposed sentiment-aware metrics compared to the frequency-based citation metrics.

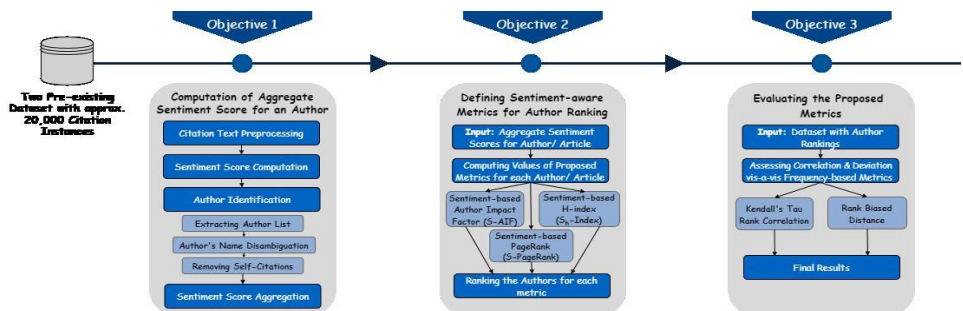

Figure 2. Flowchart detailing the steps involved in the proposed method.

## 5. Experiments and Results

In the current section, we present the experimentation performed and the results derived thereof. We have divided the same into three subsections, corresponding to our three objectives:
1. Computation of the aggregate sentiment score for an author.
2. Proposal for sentiment-aware metrics for evaluating the authors of scholarly articles.
3. Evaluation of the proposed metrics for ranking authors.

### 5.1. Objective 1: Computation of Aggregate Sentiment Score for an Author

The process starts by preprocessing the citation sentences (Section 5.1.1) following which we utilized SentiWordNet [6] to compute the sentiment scores for each citation sentence.

SentiWordNet is a specialized version of WordNet [10], a lexical database for the English language. WordNet provides definitions based on the part of speech and synsets- groups of synonyms conveying the same concept. Notably, SentiWordNet goes beyond WordNet, offering numerical scores for each synset.

SentiWordNet automatically annotates the synsets of WordNet and marks a synset (say, syn) with a three-dimensional value ($syn_+$, $syn_-$, $syn_=$) to indicate its degree of "positivity", "negativity", and "neutrality", The scores for each synset are distributed in a manner that ensures the cumulative sum equals 1, as depicted in the following Equation 1:

$$syn_+ + syn_- + syn_= = 1, \ 0 \leq syn_+, syn_-, syn_= \leq 1 \quad (1)$$

The proposed method provides a quantitative measure of sentiment for a citation sentence, based on the total sentiment score of its constituent words. For computing the score of a citation sentence, the sentiment scores of each synset in a citation sentence are added. Figure 3 shows an example for calculating the sentiment score of a citation sentence.

A cited article may be referenced in the citing article once or multiple times, each instance being termed as a citation sentence. Usually, each instance of a citation in a citing article is evaluated separately for its polarity. A function of the individual citation sentiments can be computed to ascertain the overall polarity [12]. We evaluated sentiment scores for individual citation sentences and added them to establish the overall sentiment score related to each cited article.

After computing the scores for each cited article, the subsequent step was to derive the scores for the authors (Section 5.1.2). This entailed identifying the unique authors in the dataset and removing self-citations before aggregating the scores for each author.

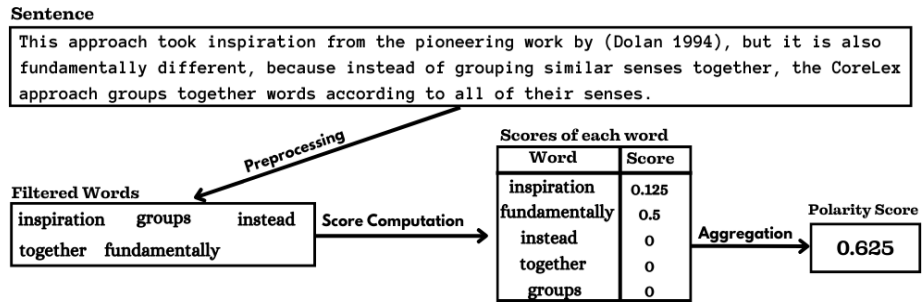

Figure 3. Example showing the computation of the sentiment score of a citation sentence.

5.1.1. Preprocessing and Sentiment-based Scoring of Citation Text

Preprocessing is a crucial component in enhancing the quality of the text data before sentiment computation. The following steps of preprocessing set the stage for a more effective and accurate sentiment score computation, aligning with the standards of scientific research:

- `Tokenization` was performed to breakdown the sentences into list of words.
- `Part-of-speech (POS) Tagging` was used to assign grammatical attributes to each token. We specifically retained lemmas categorized as adverbs, adjectives, and nouns, known for carrying sentiment-related information [19].
- `Lemmatization` simplified subsequent analysis by reducing words to their base form. Lemmatization was then done based on the POS Tags using WordNet to get the respective lemma for each word.
- We applied a `counter-based curation strategy` to remove frequently occurring lemmas that might represent domain-specific or neutral terms, minimizing potential interference with sentiment analysis [24].

`Score Computation`: We created a Python program utilizing key libraries including NLTK, Pandas, and Requests. To gather essential author information for Corpus 1, we utilized a Python script with the requests library to fetch data from the ACL Anthology, a repository renowned for housing research articles in computational linguistics. The SentiWordNet functionality of the NLTK library was employed to compute sentiment scores for individual citation sentences. It is noted that a citation sentence comprises several lemmas and multiple citation sentences may exist corresponding to a cited article in the text of a single citing article.

Consider a cited article, say 'A' mentioned $m$ times in a citing article, say 'B'. Assume that the $j^{th}$ citation sentence identified $n_j$ lemmas where $j \in [1, m]$. For the $i^{th}$ lemma in the $j^{th}$ citation sentence (say, $syn^i(j)$, $i \in [1, n_j]$) with $syn^i_+(j)$, $syn^i_-(j)$, and $syn^i_=(j)$ as positivity, negativity, and neutrality scores, the total sentiment score for the cited article 'A' w.r.t the citing article 'B' is computed as presented in Equation 2:

$$\text{Total Sentiment Score} = \sum_{j=1}^{m}\sum_{i=1}^{n}(syn_+^i(j) - syn_-^i(j)) \quad (2)$$

That is, the positive and negative sentiment values are given a weight of "1" and "-1" respectively. The neutrality score is given a weight of "0". This is done to ensure that even a small amount of sentiment score is registered in the "Total Sentiment Score". Since the scientific text mostly contains neutral words, giving a non-zero weight to the neutrality score can overshadow the positive or negative opinion of the citing author.

Additionally, using the above weights, a neutral synset will obtain a "Total Sentiment Score" of 0. Therefore, for an article cited without sentiment implication, sentiment-enhanced and traditional frequency-based metrics will lead to a similar evaluation.

### 5.1.2. Author Identification and Score Aggregation

Our primary emphasis revolved around examining author ranking based on sentiment. To achieve this, we utilized the distinct article IDs within the datasets, subsequently extracting the authors associated with each article.

**Extracting Author List:** Our objective was to automate the compilation of a list of authors for each distinct article in our dataset. To accomplish this for Corpus 1, we dynamically constructed the URL for each article by adding its unique identifier to the ACL Anthology's base URL [2] and retrieved the webpage. After successfully fetching the webpage, we employed regular expressions to parse the HTML content and extract the necessary author information. In some cases, the URL was unavailable. In such instances, we conducted manual searches using the corresponding cited/citing articles, and examined the citation sentences to extract the details of these articles. Table 2 shows the extracted author names corresponding to a subset of the articles in Corpus 1.

For Corpus 2, we automated the communication with the Semantic Scholar API [18] to retrieve Author IDs for each article. In some cases where the author IDs could not be obtained via the Semantic Scholar API, we manually searched using the citation sentences. Even after the manual search, the Author IDs could not be retrieved for 65 articles, leading us to exclude them from subsequent processes.

**Author Disambiguation:** After the initial extraction of the authors' names, we considered the following potential scenarios as motivation to disambiguate the author names:

1. Two or more (slightly different) author names may refer to the same individual.
2. The same name may be attributed to two different authors.

To address the first condition in Corpus 1, the authors were organized based on their last names and an investigation of potential similarities in the spelling and

---

[2]https://arlanthology.org/

| Article | Author1 | Author2 | Author3 | Author4 |
|---|---|---|---|---|
| A00-1004 | Chen, Jiang | Nie, Jian-Yun | | |
| A00-1005 | Bagga, Amit | Strzalkowski, Tomek | Wise, G. Bowden | |
| A00-1007 | Jonsson, Arne | Dahlback, Nils | | |
| A00-1011 | Aone, Chinatsu | Ramos-Santacruz, Mila | | |
| A00-1012 | Stevenson, Mark | Gaizauskas, Robert | | |
| A00-1014 | Chu-Carroll, Jennifer | | | |
| A00-1019 | Langlais, Philippe | Foster, George | Lapalme, Guy | |
| A00-1020 | Harabagiu, Sanda | Maiorano, Steven J. | | |
| A00-1025 | Cardie, Claire | Ng, Vincent | Pierce, David | Buckley, Chris |
| A00-1026 | Rindflesch, Thomas C. | Rajan, Jayant V. | Hunter, Lawrence | |
| A00-1031 | Brants, Thorsten | | | |
| A00-1039 | Yangarber, Roman | Grishman, Ralph | Tapanainen, Pasi | |
| A00-1042 | Wacholder, Nina | Klavans, Judith L. | Evans, David K. | |
| A00-1043 | Jing, Hongyan | | | |

Table 2

The extracted author names for a subset of the articles in Corpus 1

abbreviations of adjacent names was carried out. We compiled a list of names that could be ambiguous and searched the ACL Website, specifically in the authors' profiles under the "Also published as" section of the website. A few examples of authors along with their alternate names, identified during this disambiguation process, are as below:

- Benedi, Jose-Miguel & BenedÃ, J. M.
- Biermann, Alan & Biermann, Alan W.
- Cai, Junfu & Cai, Jun Fu
- Church, Ken & Church, Kenneth & Church, Kenneth W.
- Penstein-Rose, Carolyn & Rose, Carolyn & Rose, Carolyn P. & Rose, Carolyn Penstein

Following the above step, we refined the list of authors to eliminate any ambiguities corresponding to the first condition. For the second condition, we utilized the updated author list and performed searches for each author on the ACL website, focusing on the presence of the "Other people with similar names" section of the website. Our investigation revealed that there were no instances of the second condition in Corpus 1.

In the case of Corpus 2, since the dataset allowed us to extract the Author IDs rather than their names, it obviated the need for the author disambiguation process.

Self Citations: Self-citations occur when an article cites another article with which it shares at least one author. In our study, we carefully examined the dataset to find such instances of self-citation. To do so, a Python script was developed. The script identified matching authorship in a given pair of citing and cited papers and extracted the corresponding Article IDs. For instance, 23, and 18 self-citations were found for the authors "Collins, Michael" and "Koehn, Philipp" (Corpus 1) respectively. Subsequently, a Python script was utilized to remove all instances of citations associated with the found pairing of citing and cited articles from the dataset to ensure that self-citations did not influence our further research.

| Cited Article | Citing Article | Citation Context | Sentiment Score/ Sentence | Sentiment Score/ Cited Article |
|---|---|---|---|---|
| D07-1113 | C08-1052 | As well as the sentiment expressions leading to evaluations, there are many semantic aspects to be extracted from documents which contain writers opinions, such as subjectivity (Wiebe and Mihalcea, 2006), comparative sentences (Jindal and Liu, 2006), or predictive expressions (Kim and Hovy, 2007). | -0.5 | -3 |
| | C08-1060 | Specifically, Kim and Hovy (2007) identify which political candidate is predicted to win by an opinion posted on a message board and aggregate opinions to correctly predict an election result. | -0.875 | |
| | | Opinion forecasting differs from that of opinion analysis, such as extracting opinions, evaluating sentiment, and extracting predictions (Kim and Hovy, 2007). | -1.875 | |
| | | Kim and Hovy (2007) make a similar assumption. | 0 | |
| | C08-1101 | An application of the idea of alternative targets can be seen in Kim and Hovys (2007) work on election prediction. | 0.25 | |
| | P09-1026 | Kim and Hovy (2007) predict the results of an election by analyzing forums discussing the elections. | 0 | |
| W06-0301 | C08-1103 | (2005), Kim and Hovy (2006)), source extraction (e.g. Bethard et al. | 0 | -4 |
| | | A notable exception is the work of Kim and Hovy (2006). | 0.375 | |
| | D07-1114 | In open-domain opinion extraction, some approaches use syntactic features obtained from parsed input sentences (Choi et al., 2006; Kim and Hovy, 2006), as is commonly done in semantic role labeling. | -0.625 | |
| | | Kim and Hovy (2006) proposed a method for extracting opinion holders, topics and opinion words, in which they use semantic role labeling as an intermediate step to label opinion holders and topics. | -1.875 | |
| | | Open-domain opinion extraction is another trend of research on opinion extraction, which aims to extract a wider range of opinions from such texts as newspaper articles (Yu and Hatzivassiloglou, 2003; Kim and Hovy, 2004; Wiebe et al., 2005; Choi et al., 2006). | -1.875 | |
| | W07-2072 | Kim and Hovy (2006) integrated verb information from FrameNet and incorporated it into semantic role labeling. | 0 | |

Table 3

List of citation sentences, their sentiment scores, and the sentiment Score per cited article for the Author "Kim, Soo-Min". Two author's two articles are cited seven times. The "Aggregated Sentiment Score" for the author is -7.

Computation of Aggregate Author Score: We proceed to consolidate the "Total Citing articles" and "Total Sentiment Score" for each author within our study. The aggregation process of the sentiment score involved the summation of sentiment scores linked to the citations of the author. Table 3 displays 12 citation sentences and sentiment scores directed toward two articles authored by Kim, Soo-Min. These citations originate from seven distinct citing articles. Table 3 shows the process of computing the aggregate sentiment score for the example author to clarify the process for the same.

5.1.3. Results for Objective 1

Tables 4 and 5 present the results for the articles and authors, respectively, based on the number of citing articles and the total sentiment score value. For the purpose of discussion, we have focused on the results for the top 20 articles and authors from Corpus 1.

Among the top 20 articles (Table 4), articles such as "J93-2004", "P02-1040", and "N03-1017" exhibit a rank deviation of a maximum of two positions between rankings computed based on sentiment score and citation frequency while attaining high ranks (top 5) in both. However, it is noteworthy that articles labeled "P03-1021" and "C94-2113" exhibit a substantial difference of 174 (negative sentiment) and 53 ranks (positive sentiment) respectively, between their citation-based and sentiment-based rankings.

Among the top 20 authors (Table 5), "Marcus, Mitch", "Collins, Michael", "Papineni, Kishore", "Roukos, S.", "Ward, Todd", "Zhu, Wei-Jing", "Berger, Adam L.", and "Wu, Dekai" show a rank deviation of a maximum of two positions between sentiment and citation ranking. Conversely, authors "Carletta, Jean", "Brown, Peter F.", "Mercer, Robert L.", and "Dolan, Bill" display notable discrepancies of 20 ranks or more with a maximum variation via positive sentiment of 94 ranks for "Dolan, Bill". This assessment highlights the significant influence of both positive and negative sentiments on deviations in rankings.

### 5.2. Objective 2: Sentiment-aware Metrics for Evaluating Authors

The influence of authors/articles within the scholarly community can initially be gauged by establishing a citation network, with authors/articles connecting through citations. In a citation network, an edge is directed from the citing to the cited author or article. Figure 4 shows a subset of the authors' citation network created for our analysis with nodes representing authors of scholarly articles.

The citation network analysis brings into focus the importance of articles, authors, and ideas while identifying their interconnectedness and helps pinpoint the most influential authors/articles. Moreover, the absence of citations can be a valuable hint, revealing unexplored domains or novel ideas for future research [12].

Each citation serves as a prime candidate for sentiment analysis. For the citation network, this entails assigning sentiment-based edge weights to the directed edges in the network [12]. For example, in Figure 4, the edge from author "Liu, Qun" to author "Brown, Peter F." displays the weight 3(-0.5), implying three citations and a negative sentiment score of 0.5.

As discussed in the earlier sections, after a pre-processing of citation texts in the corpora, author identification and sentiment score aggregation for articles and authors were performed. The proposed approach employed "SentiWordNet" [6] to derive sentiment scores for individual instances.

The proposed approach then establishes a pair of directed networks for the authors; one centered around citations and the other on sentiment. In the citation network, authors were represented as nodes, while the edge weights symbolized the cumulative count of citations exchanged between cited authors and citing authors.

| Article ID | # of Citing Articles | Sentiment Score | Rank by Citation | Rank by Sentiment |
| --- | --- | --- | --- | --- |
| J93-2004 | 434 | 62.125 | 1 | 1 |
| J93-2003 | 366 | 4.875 | 2 | 25 |
| P02-1040 | 303 | 34.94 | 3 | 5 |
| P03-1021 | 271 | -7.584 | 4 | 178 |
| N03-1017 | 218 | 41.25 | 5 | 3 |
| J96-1002 | 212 | 26 | 6 | 6 |
| P97-1003 | 177 | 35.625 | 7 | 4 |
| W96-0213 | 166 | 10.5 | 8 | 17 |
| P95-1026 | 151 | 10.625 | 9 | 16 |
| J97-3002 | 147 | 19.125 | 10 | 8 |
| J93-1003 | 138 | 18.25 | 11 | 10 |
| J96-2004 | 121 | 60.75 | 12 | 2 |
| J92-4003 | 117 | 2.125 | 13 | 57 |
| W95-0107 | 109 | 4.875 | 14 | 25 |
| W02-1001 | 107 | 7.875 | 15 | 20 |
| J90-1003 | 99 | 11.5 | 16 | 15 |
| P02-1053 | 94 | 22.375 | 17 | 7 |
| J93-1007 | 71 | 3.125 | 18 | 44 |
| P90-1034 | 67 | 13 | 19 | 14 |
| W02-1011 | 66 | 18.25 | 20 | 10 |
| Remaining articles in top 20 by Sentiment Score | | | | |
| C94-2113 | 9 | 18.75 | 62 | 9 |
| P04-1035 | 59 | 17.75 | 22 | 12 |
| N03-1003 | 49 | 16 | 25 | 13 |
| P06-1101 | 24 | 9.75 | 29 | 18 |
| N06-1020 | 24 | 8.625 | 29 | 19 |
| C98-2122 | 35 | 7.875 | 26 | 20 |

Table 4

Top 20 articles (Corpus 1) by number of citations and by sentiment score value

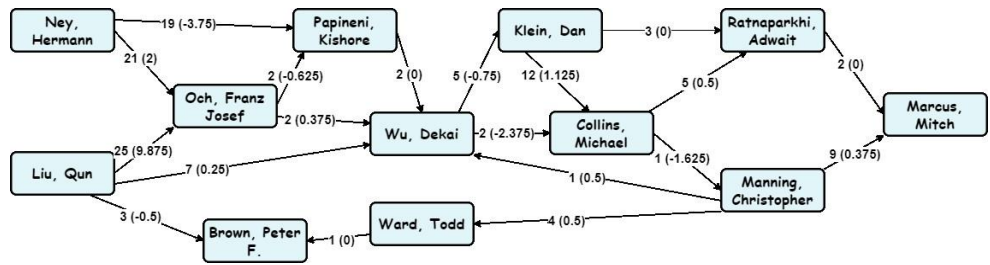

Figure 4. A subset of the author citation network (Corpus 1)

As for the sentiment-based network, it incorporated the overall sentiment score of citations to offer a fresh perspective on the ranking of the authors.

We developed three distinct alternatives for ranking based on sentiment: 1) Sentiment-based Author Impact Factor (S-AIF), 2) Sentiment-based H-index ($S_h$-index), and 3) Sentiment-based PageRank (S-PageRank) for author ranking. These alternatives are explained in the subsequent sub-sections.

5.2.1. Sentiment-based Author Impact Factor (S-AIF) and its Results

The Author Impact Factor (AIF) quantifies an author's influence in academia by measuring the citations their work receives [11]. The author's impact can be computed as the total citations (NC) to an author's publications divided by the total number of publications (NP), the formula for calculating the Author Impact Factor is expressed in Equation 3. In its calculation, AIF considers the citation frequency and does not include the notion of citation quality.

$$AIF = \frac{NC}{NP} \qquad (3)$$

In our study, we evaluate a different approach by utilizing the Total Citation Sentiment Score (S-NC), a measure of the quality of citations, departing from the traditional use of the number of citing articles. The proposed sentiment-based Author Impact Factor (S-AIF) is a score that takes into account the sentiment conveyed by citations, offering a more refined evaluation of an author's impact within academia. The formula for calculating the Sentiment-based Author Impact Factor is expressed in Equation 4. In its calculation, AIF considers the citation frequency and does not include the notion of citation quality.t

$$S - AIF = \frac{\text{S-NC}}{\text{NP}} \qquad (4)$$

Table 6 presents the results of the top 20 authors (Corpus 1) by AIF and by S-AIF.

5.2.2. Sentiment-based H-index ($S_h$-index)

The H-index [13] serves as a metric to assess the scholarly influence of an author, encompassing both the volume and significance of their publications. It signifies the count of articles (h) that have accrued no less than h citations, serving as an indicator of acknowledgment and impact. The computation methodology for H-index is outlined in Algorithm 1, providing a clear depiction of its computation process.

While the H-index enjoys widespread usage, it primarily relies on quantitative factors such as the count of citations. We introduce an innovative sentiment-driven alternative metric. In contrast to the method proposed by Zeng Ma [21], which suggests a modified version of the traditional H-index for authors using a specific

| Author Name | # Citing Articles | Sentiment Score | Rank by Citation | Rank by Sentiment |
|---|---|---|---|---|
| Della Pietra, Vincent J. | 702 | 34.625 | 1 | 15 |
| Della Pietra, Stephen | 585 | 32.5 | 2 | 16 |
| Marcus, Mitch | 543 | 67 | 3 | 1 |
| Och, Franz Josef | 512 | 39.541 | 4 | 9 |
| Brown, Peter F. | 490 | 8.625 | 5 | 34 |
| Mercer, Robert L. | 490 | 8.625 | 5 | 34 |
| Santorini, Beatrice | 434 | 62.125 | 7 | 2 |
| Marcinkiewicz, Mary Ann | 434 | 62.125 | 7 | 2 |
| Collins, Michael | 334 | 47.75 | 9 | 7 |
| Papineni, Kishore | 303 | 34.94 | 10 | 11 |
| Roukos, S. | 303 | 34.94 | 10 | 11 |
| Ward, Todd | 303 | 34.94 | 10 | 11 |
| Zhu, Wei-Jing | 303 | 34.94 | 10 | 11 |
| Marcu, Daniel | 283 | 52.375 | 14 | 6 |
| Koehn, Philipp | 241 | 45.5 | 15 | 8 |
| Berger, Adam L. | 212 | 26 | 16 | 17 |
| Lee, Lillian | 181 | 53.125 | 17 | 5 |
| Ratnaparkhi, Adwait | 166 | 10.5 | 18 | 32 |
| Wu, Dekai | 162 | 23.5 | 19 | 19 |
| Yarowsky, David | 151 | 10.625 | 20 | 31 |
| Remaining authors in top 20 by Sentiment Score | | | | |
| Carletta, Jean | 121 | 60.75 | 24 | 4 |
| Pang, Bo | 132 | 37.125 | 22 | 10 |
| Turney, Peter | 103 | 23.75 | 27 | 18 |
| Dolan, Bill | 9 | 18.75 | 114 | 20 |

Table 5

Top 20 authors (Corpus 1) by citations

| Author | AIF | S-AIF | Rank by AIF | Rank by S-AIF |
|---|---|---|---|---|
| Santorini, Beatrice | 434.00 | 62.13 | 1 | 1 |
| Marcinkiewicz, Mary Ann | 434.00 | 62.13 | 1 | 1 |
| Zhu, Wei-Jing | 303.00 | 34.94 | 3 | 3 |
| Berger, Adam L. | 212.00 | 26.00 | 4 | 4 |
| Della Pietra, Vincent J. | 175.50 | 8.66 | 5 | 10 |
| Brown, Peter F. | 163.33 | 2.88 | 6 | 27 |
| Della Pietra, Stephen | 146.25 | 8.13 | 7 | 11 |
| Dunning, Ted | 138.00 | 18.25 | 8 | 5 |
| Mercer, Robert L. | 122.50 | 2.16 | 9 | 38 |
| deSouza, Peter V. | 117.00 | 2.13 | 10 | 39 |
| Ward, Todd | 75.75 | 8.74 | 11 | 9 |
| Hindle, Donald | 67.00 | 13.00 | 12 | 6 |
| Lai, Jenifer C. | 62.00 | 1.88 | 13 | 45 |
| Marcus, Mitch | 60.33 | 7.44 | 14 | 13 |
| Cutting, Doug | 60.00 | 3.75 | 15 | 24 |
| Pedersen, Jan | 60.00 | 3.75 | 15 | 24 |
| Sibun, Penelope | 60.00 | 3.75 | 15 | 24 |
| Hanks, Patrick | 49.50 | 5.75 | 18 | 15 |
| Papineni, Kishore | 43.29 | 4.99 | 19 | 18 |
| Smadja, Frank | 35.50 | 1.56 | 20 | 54 |
| Remaining authors in top 20 by S-AIF | | | | |
| Carletta, Jean | 24.20 | 12.15 | 26 | 7 |
| Vaithyanathan, Shivakumar | 33.00 | 9.13 | 24 | 8 |
| Turney, Peter | 34.33 | 7.92 | 21 | 12 |
| Pang, Bo | 22.00 | 6.19 | 28 | 14 |
| Popat, Ashok C. | 21.00 | 5.38 | 29 | 16 |
| Dean, Jeffrey | 21.00 | 5.38 | 29 | 16 |
| Goldstein, Jade | 8.00 | 4.38 | 52 | 19 |
| Kantrowitz, Mark | 8.00 | 4.38 | 52 | 19 |

Table 6

Top 20 authors (Corpus 1) by AIF

---

**Algorithm 1** Computation of the H-index

**Require:** NP: Number of Publications, Citations: Array of publication's citations, sorted in descending order
$h \leftarrow 0$, $index \leftarrow 1$
**while** $index \leq$ NP **do**
    **if** Citations[index] $\geq$ index **then**
        $h \leftarrow index$
        $index \leftarrow index + 1$
    **else**
        break
    **end if**
**end while**
**Result:** H-index $\leftarrow h$

---

equation, our approach centers on evaluating the collective sentiment expressed within citations for each article.

The proposed metric involves examining the articles(s) published by a particular author, arranged in descending order based on their "total sentiment score". This examination provides a meaningful gauge of an author's work credibility and impact. An author is said to have a sentiment-based H-index (called $S_h$-index) score of 's' when each of the top 's' articles, ranked by total sentiment score, receives a score of at least 's'. The computational procedure for deriving the $S_h$-index is elaborated upon in Algorithm 2.

**Algorithm 2** Calculation of the sentiment-based H-index ($S_h$-index)
---
**Require:** NP: Number of Publications, Sentiment: Array of total sentiment score for publications, sorted in descending order
$s \leftarrow 0$, $index \leftarrow 1$
while $index \leq$ NP do
    if Sentiment [index] $\geq$ index then
        $s \leftarrow index$
        $index \leftarrow index + 1$
    else
        break
    end if
end while
**Result:** $S_h$-index $\leftarrow s$
---

Table 8 shows the author rankings based on H-index and $S_h$-index for the top 20 authors.

### 5.2.3. Metric based on PageRank and Sentiment Score (S-PageRank)

PageRank [8] is a graph-centered ranking algorithm that evaluates the significance of a vertex in a graph by factoring in its incoming and outgoing links [28]. In the field of author ranking, using the author citation network, a PageRank-based citation metric can gauge the author's prestige.

In the present work, we investigate author evaluation by combining the sentiment score with the PageRank approach. We present an evaluation of authors' prestige based on citation frequency vs. the sentiment score of citations.

PageRank was proposed by Google for its search engine to rank websites by considering the quantitative (frequency) and qualitative aspects of web links. The idea was to determine higher prestige web pages. The original algorithm proposed a damping factor, a value between 0 and 1, as the likelihood of a web link on the current page being picked. The choice of damping factor significantly impacts the behavior of the PageRank algorithm, prompting a closer investigation into its consequences.

Importantly, previous research [7] indicates that the optimal damping factor range varies depending on the intended purpose. Notably, the range of 0.8 to 0.9 aligns with the concept of "Weak Rank". This alignment proves valuable in scenarios prioritizing relevant results, such as web search engines. In such cases, avoiding false negatives is more important than preventing false positives. In the web search example, a false negative being a relevant page not being ranked and a false positive being an irrelevant page being ranked.

Conversely, the emphasis on "Strong Rank", achieved with damping factors between 0.5 and 0.6, is advantageous in trust and reputation systems. These systems aim to minimize false positives, making them well-suited for evaluating trustworthy items where it is crucial to avoid ranking untrustworthy items highly [7]. For instance, in academic contexts, it's particularly important to avoid false positives (untrustworthy articles being ranked highly). The aforementioned insights suggest that it is advisable to prioritize PageRank with a damping factor of 0.55 in the field of scholarly articles, given its strong reliance on trust and reputation.

We empirically tested and analyzed the rankings generated for damping factor (d) values of 0.85 and 0.55. For d=0.85, change in rankings for a few authors appeared harder to justify. For instance, upon analyzing the author "Carletta, Jean," it was observed that she was cited in a total of 124 articles and her overall sentiment score amounted to 60. Overall, she received 74 positive citations, 21 negative citations, and the rest were neutral. Citation frequency-based PageRank placed her at rank 11. Surprisingly, despite a majority of positive citations, her rank by sentiment-PageRank positioned her notably lower at 3044. In contrast, when employing a damping factor of 0.55, these ranks adjusted to 9 and 30, respectively.

It is noteworthy that Ghosh and Shah [12] delved into PageRank-based indices for articles but lacked mention of the damping factor, a critical parameter explored by us across two distinct values in our research. Additionally, we are investigating the PageRank-based metric for authors.

The results presented in Table 7 highlight the top 20 authors, ranked based on both PageRank score and on Sentiment-based PageRank (S-PageRank) score.

### 5.3. Objective 3: Evaluation of the Proposed Metrics for Ranking Authors

Our method for assessing the proposed sentiment-based metrics involves comparing them with their quantitative counterparts. As each indexing approach yields a unique sequence of articles in the collection, it's crucial to determine whether substantial divergences exist between these ordered lists. We employ two approaches to measure the degree of similarity or difference between two ranked lists:

1. Kendall's Tau Rank Correlation
2. Rank Biased Distance (RBD)

#### 5.3.1. Kendall's Tau Rank Correlation

Kendall's Tau Rank Correlation evaluates the degree of association or correlation between two sets of data by measuring the probability that two items retain their order in both lists. Positive values indicate a direct monotonic relationship, while negative values suggest an inverse relationship. Correlation coefficients with absolute values greater than 0.7 signify a strong level of correlation between variables, between 0.5 to 0.7 suggest a moderate level of correlation, and absolute values less than 0.5 indicate a weak correlation. The p-value remains a pivotal indicator of the statistical significance of these correlations, enabling us to evaluate the robustness of our findings. One remarkable feature of Kendall's Tau is its ability to handle ties effectively as compared to Spearman's Correlation Coefficient [2].

When analyzing the results for Corpus 1, we observe that Kendall's Tau ($\tau$) values consistently lie within the range of 0.35 to 0.5 for variables such as Article/Author

| Author | PageRank ($\times 10^{-2}$) | S-PageRank ($\times 10^{-2}$) | Rank By PageRank | Rank By S-PageRank |
|---|---|---|---|---|
| Della Pietra, Vincent J. | 2.73 | 3.42 | 1 | 1 |
| Della Pietra, Stephen | 2.29 | 2.52 | 2 | 4 |
| Marcus, Mitch | 2.13 | 1.87 | 3 | 5 |
| Collins, Michael | 1.87 | 0.918 | 4 | 15 |
| Brown, Peter F. | 1.76 | 2.87 | 5 | 2 |
| Mercer, Robert L. | 1.76 | 2.87 | 5 | 2 |
| Santorini, Beatrice | 1.59 | 1.31 | 7 | 9 |
| Marcinkiewicz, Mary Ann | 1.59 | 1.31 | 7 | 9 |
| Carletta, Jean | 1.43 | 0.418 | 9 | 30 |
| Yarowsky, David | 1.34 | 1.49 | 10 | 7 |
| Och, Franz Josef | 1.26 | 1.31 | 11 | 8 |
| Ratnaparkhi, Adwait | 1.16 | 1.02 | 12 | 12 |
| Dunning, Ted | 1.08 | 1.14 | 13 | 11 |
| Berger, Adam L. | 0.989 | 0.576 | 14 | 25 |
| Lee, Lillian | 0.781 | 0.141 | 15 | 52 |
| Turney, Peter | 0.744 | -0.209 | 16 | 3033 |
| Roukos, S. | 0.727 | 0.895 | 17 | 16 |
| Papineni, Kishore | 0.727 | 0.895 | 17 | 16 |
| Ward, Todd | 0.727 | 0.895 | 17 | 16 |
| Zhu, Wei-Jing | 0.727 | 0.895 | 17 | 16 |
| Remaining articles in top 20 by S-PageRank | | | | |
| Koehn, Philipp | 0.548 | 1.55 | 26 | 6 |
| Lai, Jenifer C. | 0.480 | 0.942 | 29 | 13 |
| deSouza, Peter V. | 0.464 | 0.924 | 31 | 14 |

Table 7

Top 20 authors (Corpus 1) by PageRank

| Author | H-index | $S_h$-index | Rank by H-index | Rank by $S_h$-index |
|---|---|---|---|---|
| Klein, Dan | 6 | 1 | 1 | 30 |
| Marcu, Daniel | 5 | 3 | 2 | 1 |
| Collins, Michael | 5 | 3 | 2 | 1 |
| Della Pietra, Vincent J. | 4 | 3 | 4 | 1 |
| Lee, Lillian | 4 | 3 | 4 | 1 |
| Manning, Christopher | 4 | 1 | 4 | 30 |
| Turney, Peter | 3 | 2 | 7 | 5 |
| Gildea, Daniel | 3 | 1 | 7 | 30 |
| Och, Franz Josef | 3 | 2 | 7 | 5 |
| McDonald, Ryan | 3 | 2 | 7 | 5 |
| Johnson, Mark | 3 | 2 | 7 | 5 |
| Blitzer, John | 3 | 2 | 7 | 5 |
| Pereira, Fernando | 3 | 2 | 7 | 5 |
| Hovy, Eduard | 3 | 1 | 7 | 30 |
| Ng, Andrew | 3 | 2 | 7 | 5 |
| Daume III, Hal | 3 | 2 | 7 | 5 |
| Brown, Peter F. | 3 | 2 | 7 | 5 |
| Della Pietra, Stephen | 3 | 2 | 7 | 5 |
| Mercer, Robert L. | 3 | 2 | 7 | 5 |
| Knight, Kevin | 3 | 2 | 7 | 5 |
| Remaining articles in top 20 by $S_h$-index | | | | |
| Wu, Dekai | 2 | 2 | 22 | 5 |
| Smith, David | 2 | 2 | 22 | 5 |
| Brants, Thorsten | 2 | 2 | 22 | 5 |

Table 8

Top 20 authors (Corpus 1) by H-index

Number of Citations & Sentiment Scores, AIF & S-AIF, H-index & $S_h$-index. This range signifies a weak monotonic association and a significant deviation in the author rankings for these proposed metrics when compared to their respective quantitative counterparts. However, in the case of Author PageRank (d = 0.55), Kendall's Tau ($\tau$) coefficient is relatively higher at 0.588, indicating a moderately strong relationship. It's worth noting that the findings are statistically significant at the 1% significance level.

In the context of Corpus 2, it becomes evident that Kendall's Tau ($\tau$) for all cases except the PageRank-based approach shows a weakly positive correlation. Although Kendall's Tau Coefficient for AIF & S-AIF and H & $S_h$-index remain statistically significant, the p-values for Author and Article Scores stand at 0.38 and 0.73, respectively, suggesting a lack of reliability in these cases.

In summary, sentiment awareness causes a higher disruption in frequency-based metrics as compared to PageRank-based metrics for author ranking. Table 9 displays

| Metrics | Corpus 1 | | | Corpus 2 | | |
|---|---|---|---|---|---|---|
| | Kendall($\tau$) | p-value | RBD | Kendall($\tau$) | p-value | RBD |
| Article's Number of Citations & Sentiment Score | 0.419 | $1.19 \times 10^{-15}$ | 0.385 | -0.010 | 0.408 | 0.999 |
| Author's Number of Citations & Sentiment Score | 0.433 | $1.26 \times 10^{-26}$ | 0.601 | 0.002 | 0.74 | 0.999 |
| AIF & S-AIF | 0.397 | $1.064 \times 10^{-26}$ | 0.370 | 0.014 | 0.0089 | 0.999 |
| H-index & $S_h$-index | 0.494 | $5.42 \times 10^{-20}$ | 0.314 | 0.017 | 0.0094 | 0.913 |
| PageRank & S-PageRank (d = 0.55) | 0.588 | $2.03 \times 10^{-253}$ | 0.285 | 0.795 | 0.00 | 0.282 |

Table 9

Kendall's Tau, p-value, and Rank Biased Distance (RBD) for the proposed sentiment-based metrics compared to their quantitative counterparts the computed Kendall's coefficients for the proposed indices in comparison to their quantitative counterparts for both datasets.

To understand the above results better, we studied the characteristics of both corpora. Diverging from the concentrated citation pattern observed in Corpus 1 (with a total of 195 cited articles), Corpus 2 exhibits a distinctive arrangement. An intriguing observation emerges within Corpus 2 that a substantial 6,405 articles, out of 11,020 instances, are referenced by just a single article each. This distribution highlights that a notable proportion of articles in Corpus 2 have garnered minimal citations. Consequently, these articles tend to be clustered together in terms of rank, a consequence of their shared attribute of possessing only one citation.

Equally noteworthy, this trend extends to the realm of authors in Corpus 2. Among the 24,024 cited authors, a significant 22,890 authors also have just single citations. As a result, these instances end up occupying identical ranks in relation to the number of citing papers.

The sentiment score introduces an element of diversity due to its reliance on the content of the citation sentence, rather than being influenced solely by the quantity of citations. Notably, these metrics serve as effective differentiators, functioning as tiebreakers for example in our case where a large number of authors of the corpus exhibits only one citation, as is further shown in Section 6.

5.3.2. Rank Biased Distance (RBD)

Rank Biased Distance (RBD) [35] is a metric to measure the extent of deviation between the two ranked lists, giving more weight to the elements at higher ranks as compared to those at the lower ranks. While Kendall's Tau rank correlation coefficient does not distinguish between ranks based on position as long as the relative order is preserved, in the context of ranked lists, it is often considered that items ranked higher carry greater importance than those ranked lower. In other words, a rank switch at the top of a ranked list should lower the similarity score compared to a rank switch at the bottom of the list. RBD is sensitive to the switching of the ranks of articles at the top of the ranked list and is computed using the Equation 5 presented below:

$$RBD_k = (1-p) \sum_{d=1}^{k} \frac{S_d \Delta T_d}{2d} \cdot p^{d-1} \qquad (5)$$

In the formula for RBD presented above, S and T are the two ranked lists being compared, each containing k elements. Δ symbolizes the deviation between the two ranked lists. p is a weighing factor with a value in the range [0, 1]. A lower value of p assigns more weight to higher-ranked elements. Since a higher value of p evenly distributes the weight among all the elements of the ranked list, a value of 0.9 was adopted for p, which is also in line with recommendations from the original work [35]. Next, we interpret the Rank Biased Distance (RBD) results, tabulated in Table 9.

In Corpus 1, for most indices, the RBD value is lower than 0.4, suggesting substantial but reasonable ranking shifts when comparing sentiment-based and citation-based metrics. Notably, the RBD value for the Author's Number of Citations vs. Sentiment Score exhibits a relatively higher deviation of 0.601, signifying significant disparities.

In contrast, in Corpus 2, citation-based rankings tend to cluster around specific values, while sentiment-based rankings exhibit broader variation based on sentiment scores. This variability likely contributes to the increased deviations for most indices with RBD scores close to 1. Additionally, the PageRank-based index in Corpus 2 shows a lower deviation since it already has an embedded consideration of prestige.

## 6. Discussion

### 6.1. Discussion on Results for Objective 1

Our first objective dealt with the computation of aggregate sentiment scores for authors and its experimental results are tabulated in Tables 4 and 5. To better explain the results, we analyzed the article "P03-1021" (Och and Ney, 2003) (Table 4) that exhibited maximum rank deviation. It is observed that many of the citation sentences point out the limitations of the article "P03-1021" and highlight the ongoing advancements in the field. Following are two such examples of citation sentences:

"The most widely applied training procedure for statistical machine translation IBM model 4 (Brown et al., 1993) unsupervised training followed by post-processing with symmetrization heuristics (Och and Ney, 2003) yields low-quality word alignments."

"Unlike minimum error rate training (Och and Ney, 2003), our system can exploit large numbers of specific features in the same manner as static reranking systems (Shen et al., 2004; Och et al., 2004)."

The above citation sentences express a negative sentiment for the article "P03-1021" and talk about the progressions beyond it, thereby validating the rank deviation.

In addition, we consider the following citation sentence citing the article "W05-0904" (Liu and Gildea, 2005). The provided sentence exemplifies a scenario in which criticism is succeeded by a subtle acknowledgment of merit.

"Our method, extending this line of research with the use of labelled LFG dependencies, partial matching, and n-best parses, allows us to considerably outperform Liu and Gildea's (2005) highest correlations with human judgement (they report 0.144 for the correlation with human fluency judgement, 0.202 for the correlation with human overall judgement), although it has to be kept in mind that such comparison is only tentative, as their correlation is calculated on a different test set."

The above sentence conveys a negative opinion about the cited article "W05-0904." However, the tone of the sentence appears positive in that it praises the citing article's work. Analyzing the sentiment score, the given sentence has an overall sentiment score of "-0.75". Meanwhile, the citation frequency-based and sentiment score-based evaluations resulted in 48 and 151, respectively, indicating that the criticism is noted, despite the positive tone of the sentence, while manifesting the cited article's rank.

## 6.2. Discussion on Results for Objective 2

In the present section, we discuss the results for each of the three developed sentiment-enhanced citation metrics, namely: S-AIF: Sentiment-based Author Impact Factor, $S_h$-index: Sentiment-based H-index, and S-PageRank: Sentiment-based PageRank.

### 6.2.1. S-AIF

The results for S-AIF are tabulated in Table 6. An analysis of the rankings given in Table 6 emphasizes the interaction between citation-based and sentiment-based impact by highlighting cases in which AIF and S-AIF rankings don't demonstrate substantial changes and situations in which they have significant variations.

As an illustration, writers such as "Santorini, Beatrice" and "Marcinkiewicz, Mary Ann" (Table 6) maintain a similar position in both metrics, however "Smadja, Frank" and "Mercer, Robert L." possess a substantially higher rank by AIF (20 & 9) compared to their rank by S-AIF (54 & 38).

Analyzing the author "Santorini, Beatrice" (Table 6), we discovered that they have only one article titled "J93-2004." This article has been cited in 434 other articles, with a total of 626 citation sentences. Among these sentences, 264 received a sentiment score of "0," 246 received a score higher than 0, and 116 received a negative sentiment score. The prevalence of neutral and positive sentences has resulted in a high S-AIF score of 62.13. The high AIF as well as the high S-AIF indicates indicates that the author's work has received an overall positive recognition, resulting in a strong ranking in both metrics.

Examining the author "Mercer, Robert L." (Table 6) with an AIF score of 122.50, we found that he has authored four publications. These articles have been cited in 490 other articles, generating 918 citation sentences. Out of these sentences, 408 received a sentiment score of "0," 270 had a score above 0, and 240 had a negative sentiment score. Although there are more neutral sentences, the number of positive and negative sentences is roughly balanced, resulting in a slightly positive overall sentiment score of 2.16. The presence of a slightly positive sentiment score as compared to a moderately strong AIF score explains the observed deviation. This suggests that Robert's work might have garnered a significant number of citations, contributing to a high AIF rank, but the sentiment analysis reveals a lower emotional impact associated with their work.

### 6.2.2. $S_h$-index

Table 8 shows the results for $S_h$-index. It can be observed that "Marcu, Daniel" and "Collins, Michael" (Table 8 ) maintain stable positions across both metrics, while "Klein, Dan" and "Manning, Christopher" hold notably higher H-index ranks (1 & 4) compared to their $S_h$-index ranks (30).

Examining all seven articles authored by "Klein, Dan" (Table 8 ), we observed that six of these articles were cited by six or more other articles, resulting in an h-index of 6. However, when evaluating sentiment scores, only four out of the seven articles received an overall positive score, with just one article achieving a score greater than 1. Consequently, the $S_h$-index was determined to be one. That is, the works of "Klein, Dan" garnered citations, boosting his H-index, but also received lower sentiment scores, suggesting a less intense emotional impact associated with his articles.

In summary, Table 8 highlights the interplay between citation-based influence and sentiment-based H-index rankings. It showcases scenarios with consistent H-index and $S_h$-index rankings as well as those with significant disparities.

### 6.2.3. S-PageRank

The results for S-PageRank are presented in Table 7. Upon analyzing the data, it becomes evident that certain authors, such as "Della Pietra, Vincent J.", "Della Pietra, Stephen", and "Marcus, Mitch" (Table 7) exhibit minimal disparity between these two ranking methods. Conversely, the author "Turney, Peter" demonstrates a substantial discrepancy of 3017 ranks between the two rankings.

Furthermore, within the context of the PageRank algorithm, it becomes apparent that authors lacking any citations within the considered dataset would receive a popularity score of zero for conventional metrics like the H-index and impact factor. However, the PageRank-based metrics could potentially assign a non-zero score to such authors resulting in interesting results such as an uncited author ranked higher than a cited author.

We consider an uncited author "Liu, Hao" who received a PageRank score of 0.00018 and an S-PageRank score of 0.00021. On the other hand, an author "Klein, Dan" has been cited in 86 articles with a PageRank score of 0.00372 and an S-PageRank score of -0.00995. Upon analyzing "Klein, Dan," we found that 53 authors cited him negatively, 51 cited him positively, and 36 cited him with a sentiment score of 0 resulting in him receiving a lower S-PageRank based rank than the uncited author.

### 6.3. Discussion on Results for Objective 3

The evaluation of the proposed metrics uncovers valuable insights (Table 9). In Corpus 1, Kendall's coefficient reveals a weak and positive link between citation-based metrics and the proposed sentiment-based metrics. Notably, this relationship triggers substantial shifts in author rankings for articles with pronounced (positive or negative) sentiment. For instance, consider the author "Banerjee, Satanjeev" with 55 citations from 3 publications and a cumulative sentiment score of -1.875 showing an overall negative sentiment of the citations. His AIF score of 18.33 contrasts sharply with an S-AIF of -0.625, leading to a significant rank difference of 262 between AIF (Rank 32) and S-AIF (Rank 294). A similar trend of high deviation is observed for H-index vs. $S_h$-index (rank difference of 167) and S-PageRank and PageRank (rank difference of 2993). These variations underscore the significant impact of (negative) sentiment on metrics for "Banerjee, Satanjeev."

In Corpus 2, Kendall's coefficient for PageRank vs. S-PageRank exhibited a robust correlation of approximately 0.82. AIF/S-AIF and H-index/$S_h$-index yielded weak but positive and significant correlation.

Notably, corpus 2 contains a large number of articles and authors with the same rank values for quantitative metrics. The data presented in Table 10 illustrates four distinct instances from corpus 2, wherein authors possess a citation count of one. Although the citation count results in an Author Impact Factor (AIF) of 1 for each case, the varying sentiment scores of the associated articles contribute to divergent Sentiment-based Author Impact Factors (S-AIF). Consequently, while the AIF rankings position all instances similarly, the S-AIF yields disparate results, prioritizing qualitative factors over frequency-based parameters. This nuanced approach proves valuable, particularly for recent research articles, by effectively distinguishing superior quality papers within the academic landscape.

For Corpus 1, it is observed that the inclusion of sentiment in evaluating the impact of authors and articles leads to at least a 28% difference between the ranking metrics. For Corpus 2, this difference is about 99% in most of the metrics but for PageRank-based metrics, the difference is the minimum at 28%.

PageRank exhibited higher Kendall's coefficients in both Corpus 1 and Corpus 2 because it incorporates prestige when assessing node importance within a network. Consequently, the correlation between PageRank and S-PageRank remained consistently strong, indicating that the addition of sentiment enhances results to a limited extent, preserving the core prestige-based attributes of PageRank. This consideration differentiates it from metrics solely focused on popularity.

### 7. Summary, Conclusions, and Future Work

In this article, we examined how sentiments influence both, individual articles and their authors. Our article's overall focus can be categorized into four main sections. Our initial step was to evaluate the sentiment score of each citation sentence. To ac-

| Cited Author's ID | Citation Text | Sentiment Score | Ranks by AIF | Ranks by S-AIF |
|---|---|---|---|---|
| 1802065 | Eyelid suture, on the other hand, produces a novel pattern of thin dark columns alternating with wide pale columns (Horton, 1984; Hendry and Jones, 1986; Crawford et al., 1989; Trusk et al., 1990, Tigges et al., 1992). | 0.5 | 11588 | 4414.5 |
| 3881197 | CDDO-Me is a synthetic triterpenoid that was under phase III clinical development for the treatment of advanced chronic kidney disease (37, 38) However, due to adverse events in the phase III clinical trial, further development of CDDO-Me was terminated (39). | -1.5 | 11588 | 23267 |
| 10616537 | Our data also showed that lesions in the head and neck region responded better to PDL therapy than in other regions; similar results were reported by several studies [16, 24Aa'€vs28]. | 0.75 | 11588 | 2681 |
| 145396957 | The recent identification of lymphatic endothelial-specific markers, such as hyaluronic acid receptor-1 (LYVE-1) [2], has greatly increased attention on how lymphangiogenesis, the growth of lymphatic vessels, is regulated in the tumor microenvironment. | 0.125 | 11588 | 8992.5 |

Table 10

Authors with a single citation but different sentiment scores. While their AIF-based rank is the same, the S-AIF-based ranking shows a marked difference.

complish this, we applied text preprocessing techniques and subsequently harnessed the SentiWordNet lexical resource for calculating the sentiment score of each citation sentence. Through this process, we derived sentiment scores for each instance within the Citation Sentiment Corpus [5] (Corpus 1) and Citation Intent Classification Dataset [9] (Corpus 2).

Our second task involved the computation of overall sentiment scores for both articles and authors. This involved a multi-step process. Initially, we extracted author information for each article. Subsequently, we implemented a self-citation filter by identifying pairs of citing and cited articles that shared at least one author in common. Finally, we aggregated the sentiment scores associated with each citation within an article to determine the "Total Sentiment Score" for that specific article. Similarly, for author sentiment scores, we aggregated the "Total Sentiment Scores" derived from all articles authored by that individual.

Our third task was to introduce sentiment-driven metrics to evaluate authors and articles, leveraging the total sentiment scores computed in the previous step. Building upon commonly used quantitative metrics like Author Impact Factor (AIF), H-index, and PageRank, we put forth their sentiment-oriented equivalents, namely S-AIF, $S_h$-index, and S-PageRank.

Finally, we assessed the performance of the proposed sentiment-driven indices in comparison to their quantitative counterparts, using Kendall's Tau correlation coefficients and associated p-values. Along with this, we also employed Rank Biased Distance (RBD) to quantify the deviation between the order of rankings.

In both datasets, for most of the metrics, a weak but monotonic relationship was observed between the frequency-based and sentiment-based metrics. Interestingly, across both corpora, the Pagerank-based indices exhibited a higher correlation, indicating that prestige-based ranking had a stronger association between citation-based and sentiment-based counterparts as compared to purely popularity-based ranking. The present study also demonstrated that a sentiment-aware index leads to at least 28% rank-based deviation in author ranks from a citation-based index.

In summary, we addressed all our research objectives successfully. We established a method for computing a composite score value for an author. The proposed sentiment-infused citation metrics revealed significant shifts in author rankings for articles that were cited with a stronger (positive or negative) sentiment.

This relationship triggers substantial shifts in author rankings for articles with pronounced (positive or negative) sentiment. Our work establishes the importance of considering the qualitative aspects of a citation while evaluating the influence of an authors' work. An evaluation of scholarly impact solely on the basis of citation frequency tends to follow the tenet of "rich becoming richer". Incorporating sentiments in the analysis of an author's impact lends more depth to the evaluation. However, we believe that more research needs to be carried out on larger data to establish the widespread use of sentiment-aware indices.

The field of Scientometrics is poised to consider the sentiments of citations during the evaluation of authors and their articles. To do the same, the researchers require a shift in the way they analyze the research output. However, to achieve this, it is necessary to make the computation and use of the proposed sentiment-based metrics easy and accessible.

Our analysis is focused on determining the significance of an article from the eye of the sentiments expressed by the citing authors. Notably, certain aspects of context and period can be reflected by the sentiment embedded in the citation. For instance, certain aged publications amass a substantial number of citations, maintaining a prominent position in scholarly rankings despite their outdated nature. While these outdated articles may initially garner positive sentiments, over time, they transition into a reference in the literature, eliciting predominantly neutral sentiments. By considering sentiments, our approach aims to provide a more nuanced evaluation, allowing newer and impactful articles of authors a chance to gain better ranking positions.

In the future, we may study the integration of factors such as the citation distribution of an article/author, and the proportion of neutral citations into the ranking of scholarly articles and authors. We also plan to develop an automated system that given an author's details should extract citation sentences from articles citing the authors' publications and compute the author's sentiment score and the value of the sentiment-based metrics.

## Ailiations

Shikha Gupta
   Assoriate Professor, Computer Srienre, Shaheed Sukhdev College of Business Studies, New Delhi, India, shikhagupta@ssrbsdu.ar.in

Animesh Kumar
   Student, Computer Srienre, Shaheed Sukhdev College of Business Studies, New Delhi, India, animesh.2l508@ssrbs.du.ar.in